\documentclass[sigconf]{acmart}

\usepackage{multirow}
\usepackage{subcaption}
\usepackage{hyperref}

\AtBeginDocument{%
  \providecommand\BibTeX{{%
    \normalfont B\kern-0.5em{\scshape i\kern-0.25em b}\kern-0.8em\TeX}}}

\setcopyright{acmlicensed}
\copyrightyear{2024}
\acmYear{2024}
\acmDOI{XXXXXXX.XXXXXXX}

\acmConference[Conference acronym 'XX]{Make sure to enter the correct
  conference title from your rights confirmation email}{June 03--05,
  2018}{Woodstock, NY}
\acmISBN{978-1-4503-XXXX-X/18/06}

\graphicspath{{./images/}}

\begin{document}

\title{EHR-Based Mobile and Web Platform for Chronic Disease Risk Prediction Using Large Language Multimodal Models}

\author{Chun-Chieh Liao$\dagger$, Wei-Ting Kuo$\dagger$, I-Hsuan Hu, Jun-En Ding, Feng Liu}
\email{cliao9@stevens.edu}
\orcid{1234-5678-9012}
\affiliation{%
  \institution{Stevens Institute of Technology}
  \streetaddress{1 Castle Point Terrace}
  \city{Hoboken}
  \state{New Jersey}
  \country{USA}
  \postcode{07030}
}

\author{Yen-Chen Shih}
\email{bnn01615@gmail.com}
\affiliation{%
  \institution{Northeastern University}
  \city{Seattle}
  \state{Washington}
  \country{USA}
}

\author{Fang-Ming Hung\textsuperscript{*}}
\email{philip@mail.femh.org.tw}
\affiliation{%
  \institution{Department of Surgical Intensive Care Unit, Far Eastern Memorial Hospital}
  \city{New Taipei City}
  \country{Taiwan}
}

\renewcommand{\shortauthors}{Trovato and Tobin, et al.}

\begin{abstract}
  Traditional diagnosis of chronic diseases involves in-person consultations with physicians to identify the disease. However, there is a lack of research focused on predicting and developing application systems using clinical notes and blood test values. We collected five years of Electronic Health Records (EHRs) from Taiwan's hospital database between 2017 and 2021 as an AI database. Furthermore, we developed an EHR-based chronic disease prediction platform utilizing Large Language Multimodal Models (LLMMs), successfully integrating with frontend web and mobile applications for prediction. This prediction platform can also connect to the hospital's backend database, providing physicians with real-time risk assessment diagnostics. The demonstration link can be found at \url{https://www.youtube.com/watch?v=oqmL9DEDFgA}.
\end{abstract}



\received{20 February 2024}  
\received[revised]{12 March 2024} 
\received[accepted]{5 June 2024}  

\maketitle

\noindent\rule{4cm}{0.4pt}

\noindent $\dagger$ These authors contributed equally.
\\
\noindent \textsuperscript{*} Corresponding Author.

\begin{figure}
\centering
\includegraphics[width=0.4\textwidth]{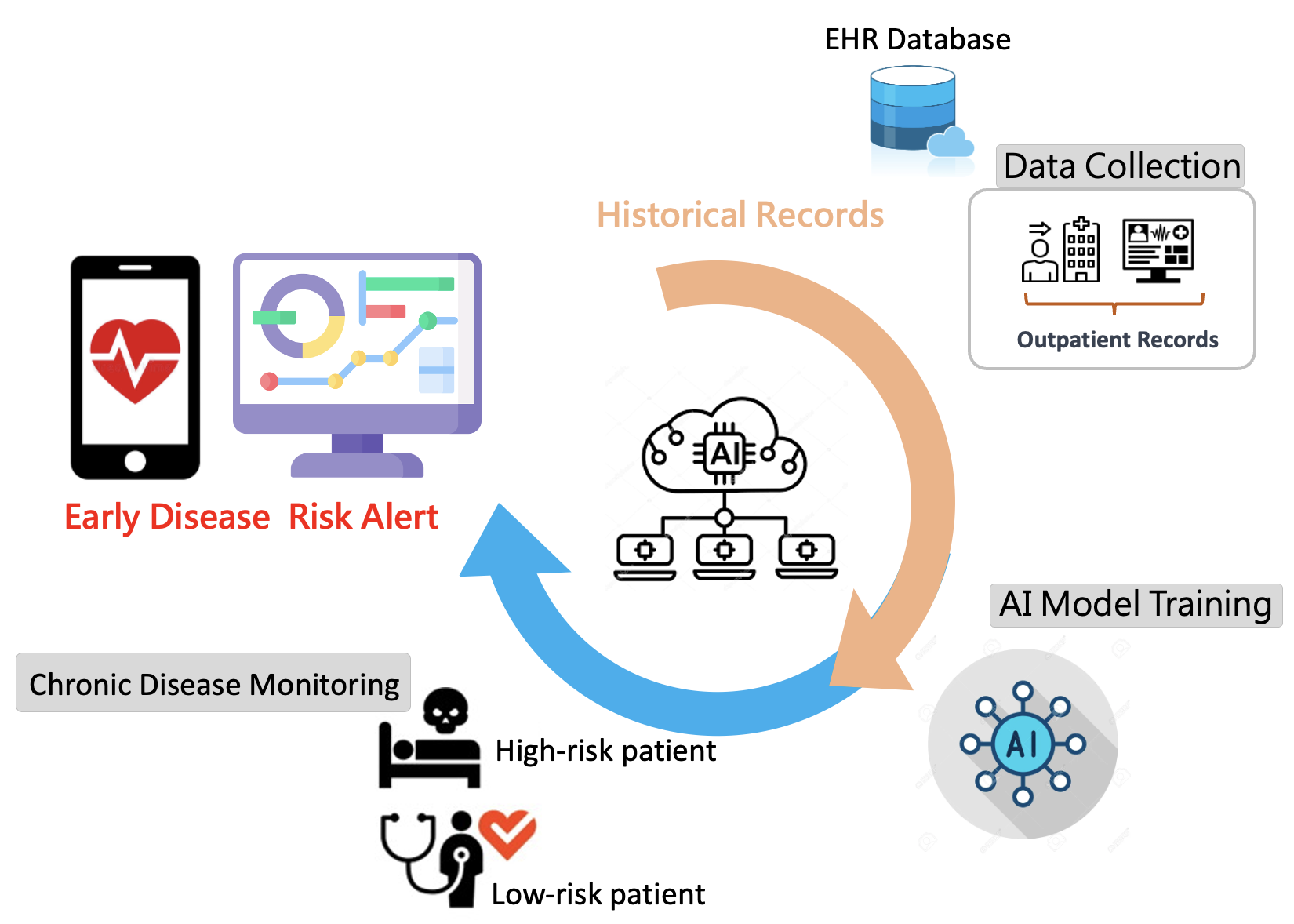}
\caption{An overview of the AI-driven disease prediction and alert system.}
\label{fig:Fig_1}
\end{figure}

\section{Introduction}

Chronic diseases such as diabetes, high blood pressure, and heart disease are all diseases of concern in many countries \cite{who-2020,chew2023global}. These chronic diseases are also associated with a high incidence of mortality \cite{global2023global,schlesinger2022prediabetes}. Traditional diagnosis of chronic diseases involves in-person consultation with a physician to identify the disease. However, this will result in a significant waste of time and medical resources. 

In the hospital diagnosis system, most patient records are stored in digital format using Electronic Health Records (EHRs), including patient clinical notes, blood test results, and pathology reports. Clinical notes are typically recorded in the database by doctors after they have seen the patient. Particularly, EHRs encompass multimodal data, including numerical values (e.g., blood test results) and categorical data (e.g., gender, age). In recent years, advancements in deep learning technology have significantly enhanced natural language processing (NLP), making it a primary focus in the research of disease classification within clinical notes . NLP techniques have demonstrated considerable potential in comprehending the contextual information embedded in medical domain sentences \cite{lee2023assessment,jiang2023health}.

In recent years, large language models have demonstrated remarkable performance in medical question answering and diagnosis, as well as in using NLP to predict various diseases \cite{Zhao-2023,yang2022large,ding2024large}. For various unstructured data types, multimodal NLP has been increasingly applied to diagnose and classify diseases by integrating clinical notes and medical images with different domain features \cite{cahan2023multimodal}. In the EHR data remote monitoring platform, a mobile system has been established for sharing lung and health data, allowing for the remote monitoring of patients' conditions \cite{genes2018smartphone}. However, there is a paucity of research focused on predicting and developing application systems using clinical notes and blood test values. In this study, we present several key contributions:

\begin{enumerate}
    \item Utilizing  large language multimodal models for predicting multiple chronic diseases using modality data.
    \item Providing an interpretable early diabetes prediction platform.
    \item Developing a multimodal health data prediction language model with real-time mobile and web-based data collection platforms.
\end{enumerate}

\section{Data Collection}

In this study, we collected five-year EHRs from the Far Eastern Memorial Hospital (FEMH) Taiwan hospital database from 2017 to 2021, including 1,420,596 clinical notes, 387,392 laboratory results, and more than 1,505 laboratory test items. The database included clinical notes and laboratory results. The study was approved by the FEMH Research Ethics Review Committee (\href{https://www.femh-irb.org/}{https://www.femh-irb.org/}) and data has been de-identified.  We first conducted data processing and physician annotation and established a comprehensive database integrated with the AI server system. Finally, we developed a complete architecture for the user interface (UI) and mobile end, as shown in Figure  \ref{fig:Fig_1}.

\subsection{Large Language Multimodal Models}

\begin{figure}
\centering
\includegraphics[width=0.5\textwidth]{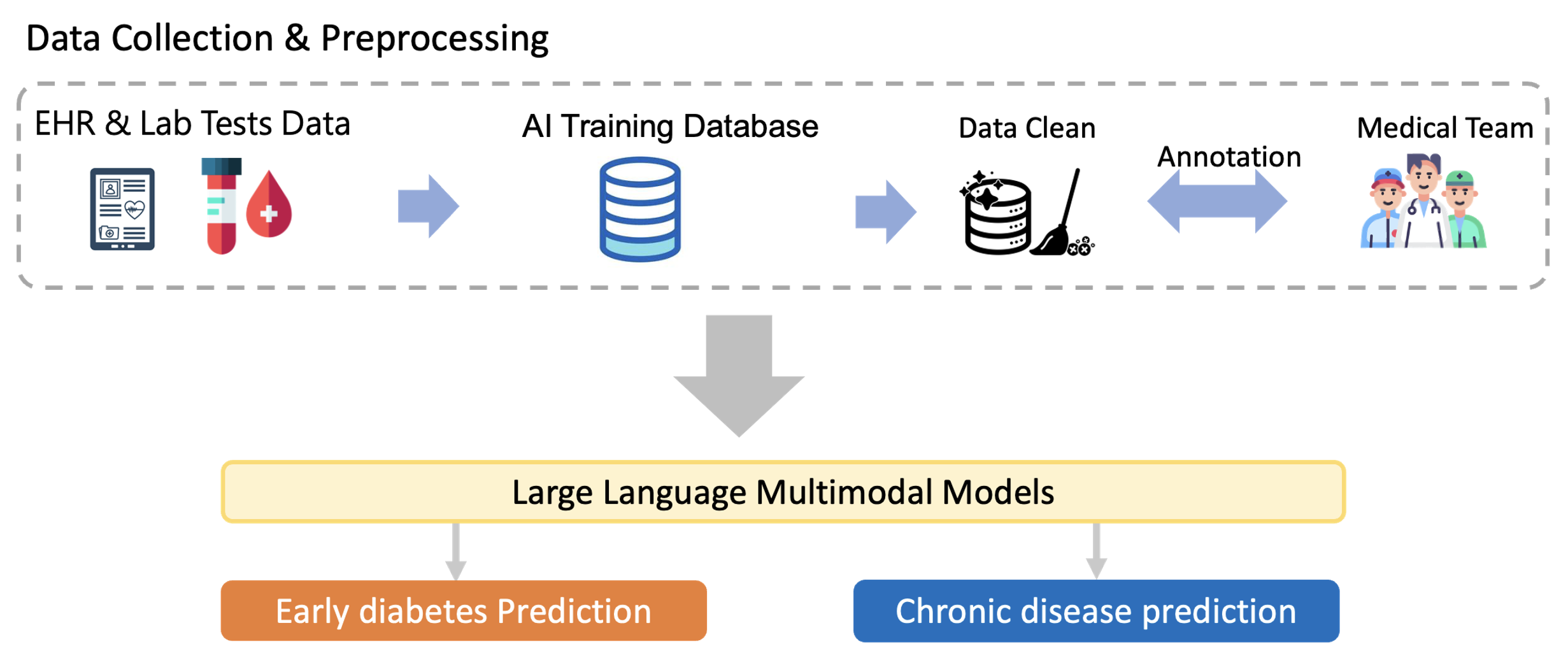}
\caption{The workflow of building AI database and LLMMs training and inference tasks.}
\label{fig:LLMMs}
\end{figure}

In our study, we utilized clinical notes and blood test data related to common chronic diseases such as diabetes, heart disease, and hypertension to conduct multimodal model training. Specifically, we employed widely used language models such as BERT \cite{devlin2018bert}, BiomedBERT \cite{gu2021domain}, Flan-T5-large-770M \cite{chung2022scaling}, and GPT-2 \cite{radford2019language} as a text feature extractor. Next, we integrate clinical notes from a single modality as input into LLMMs to extract text feature embeddings and fuse them using an attention module for the final prediction task, as shown in Figure \ref{fig:LLMMs}.

\begin{figure*}
\centering
\includegraphics[width=0.7\textwidth]{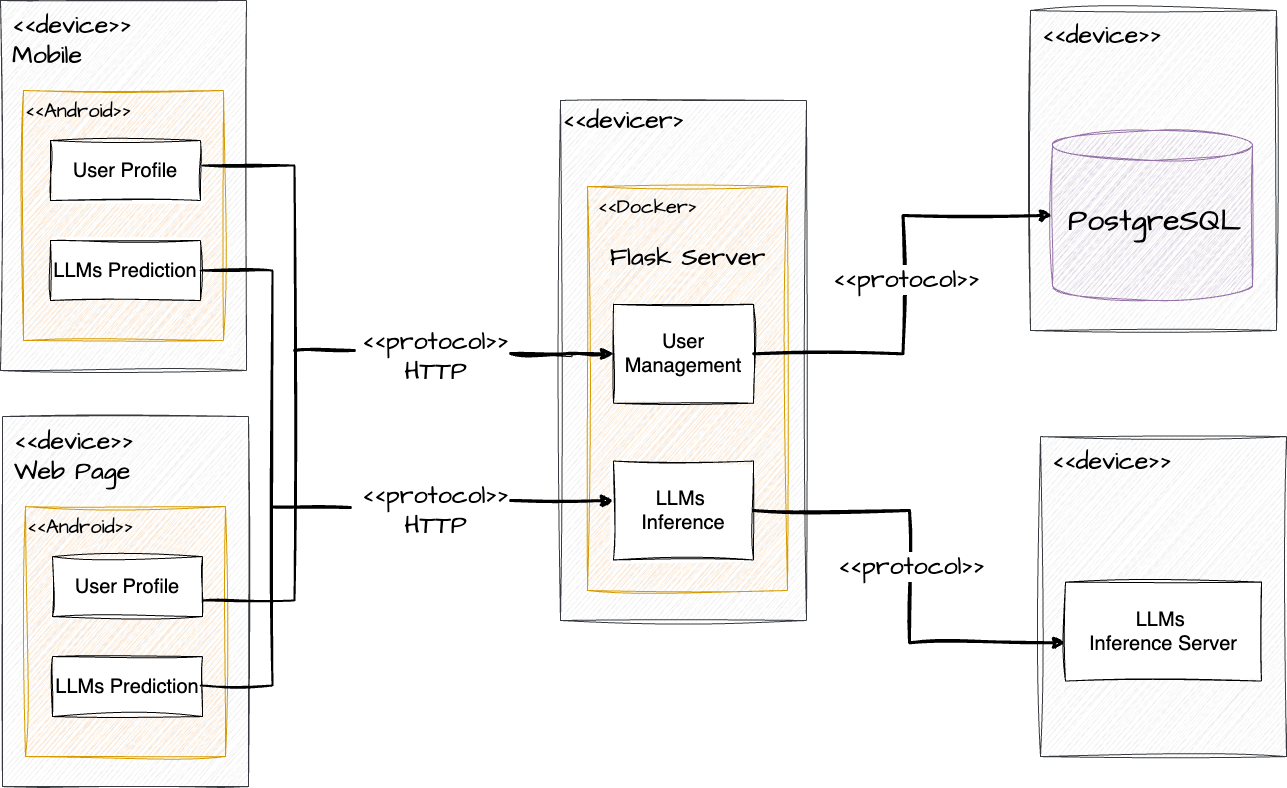}
\caption{The system design architecture that mobile and web apps interact with a Flask server for user management and LLMs inference, communicating with a PostgreSQL database and an LLMs inference server.}
\label{fig:system_design}
\end{figure*}

\subsection{Multimodality and Data Fusion}

 For the blood test data, we build a Deep Neural Network (DNN) to obtain the blood presentation. To better integrate the two modalities, we utilized a multi-head attention layer to compute the attention scores and matrices for the embeddings from both domains. Finally, fully connected layers were employed to predict multiple diseases.

\begin{figure*}[ht]
\centering
\includegraphics[width=1\textwidth]{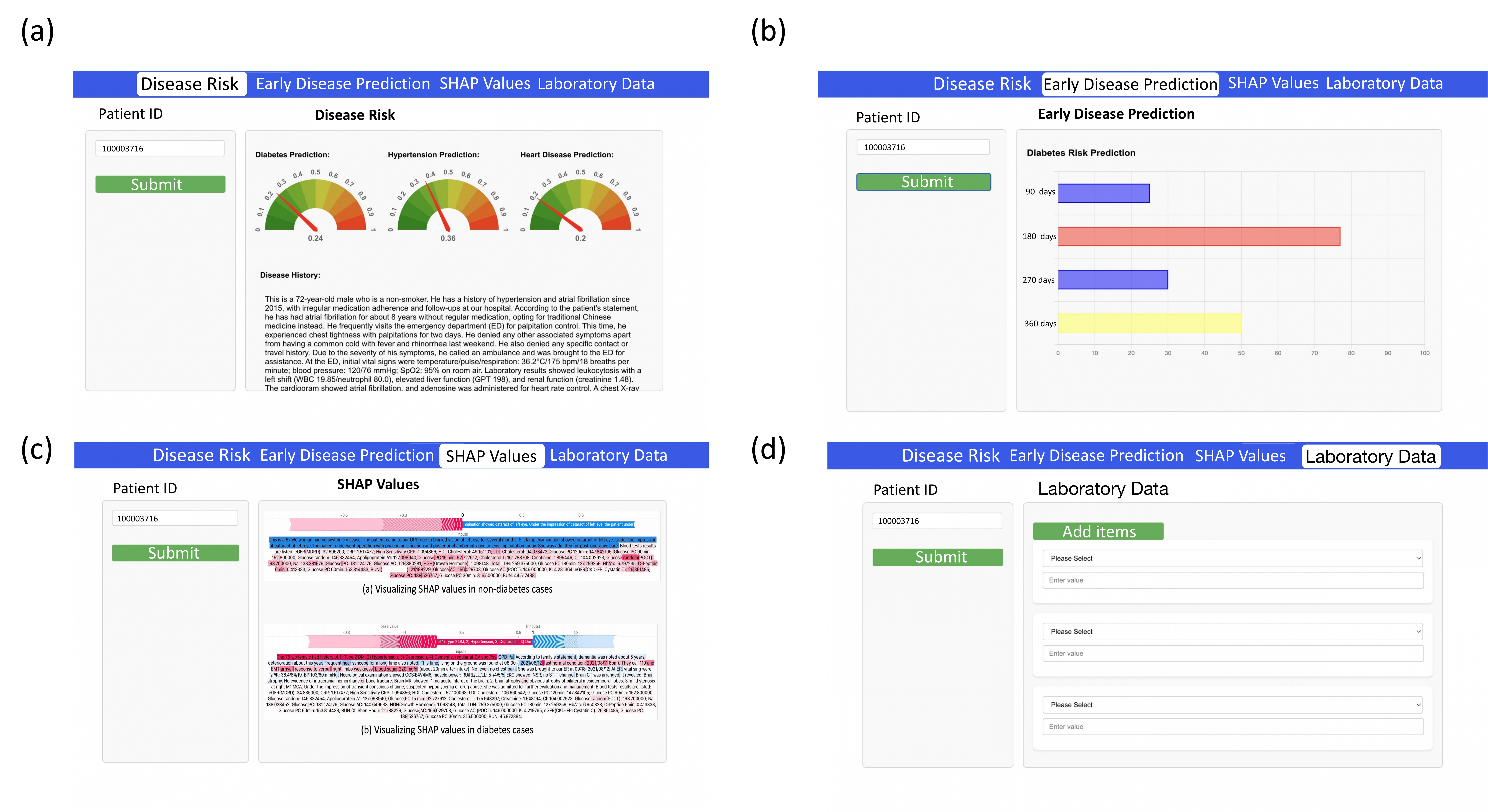}
\caption{The interface of the medical diagnostic Web system}
\label{fig:web_UI}
\end{figure*}

\begin{figure*}
\centering
\includegraphics[width=1\textwidth]{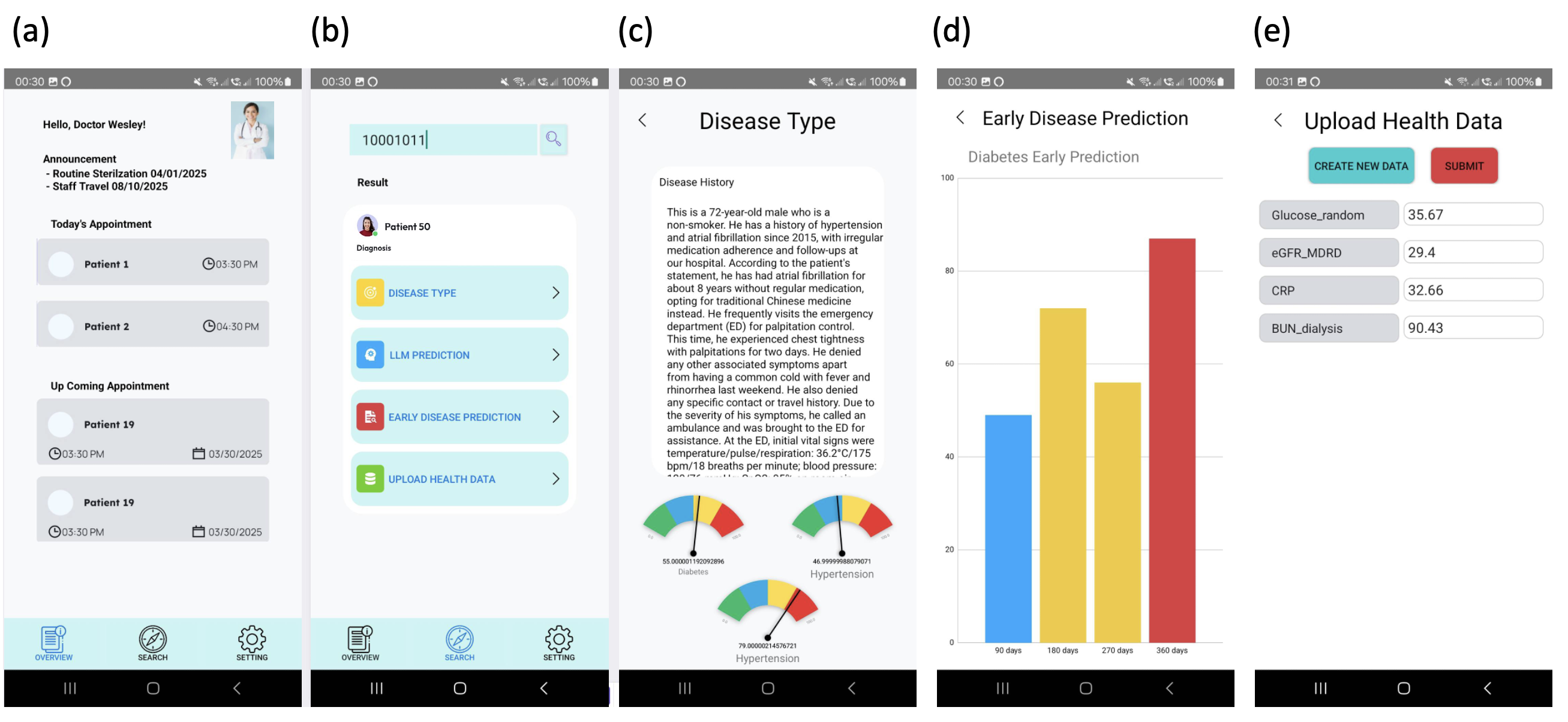}
\caption{The real-time medical appointment and diagnosis mobile platform.}
\label{fig:mobile_UI}
\end{figure*}

\subsection{Model Evaluation}

To better evaluate the unimodal performance of LLMs. Table \ref{tab:multiclass} shows that the performance of LLMMs varies depending on the positive rate of different samples, including diabetes (20.4\%), heart disease (22.57\%), hypertension (3.3\%). It is worth noting that when classifying certain specific diseases, especially those with a lower positive class, the performance of GPT-2 is not particularly well. In contrast, BiomedBERT  with prior knowledge achieve an precision of 0.35 for hypertension. In contrast, in classes with higher positive rates, such as diabetes, the combination of LLMMs's modality data with GPT-2 achieved an precisoin of 0.70, a recall of 0.71, and an F1 score of 0.70. For heart disease, GPT-2 showed a significant improvement, reaching a precision of 0.81, a recall of 0.85, and an F1 score of 0.83. Based on the experimental findings, we determined that applying distinct unimodal language models with DNN to various diseases within LLMMs generate different impacts and achieved more stable and superior performance in multiclass prediction.

\begin{table}[H]
\centering
\resizebox{\columnwidth}{!}{%
\begin{tabular}{|l|l|l|l|l|}
\hline
\multirow{2}{*}{Disease type} & \multirow{2}{*}{Models} & \multicolumn{3}{l|}{Metrics} \\ \cline{3-5} 
                              &                         & Precision & Recall & F1     \\ \hline
\multirow{5}{*}{\shortstack[l]{Hypertension\\ (n=1,230)}} 
& BERT        & 0.35      & 0.32   & 0.33   \\ 
& \textbf{BiomedBERT}  & \textbf{0.35}      & \textbf{0.29}   & \textbf{0.32}   \\ 
& Flan-T5-large-770M    & 0.29      & 0.16   & 0.20    \\ 
& GPT-2   &  0.29     & 0.21   & 0.25  \\ \hline
\multirow{5}{*}{Heart disease (n=6,929)} 
& BERT        & 0.71      & 0.76   & 0.52   \\ 
& BiomedBERT  & 0.76      & 0.75   & 0.75   \\ 
& Flan-T5-large-770M    & 0.70      & 0.78   & 0.74   \\ 
& \textbf{GPT-2}   & \textbf{0.81}     & \textbf{0.85}   & \textbf{0.83} \\ \hline
\multirow{5}{*}{Diabetes (n=7,208)} 
& BERT        & 0.66      & 0.58   & 0.62   \\ 
& BiomedBERT  & 0.63      & 0.72   & 0.67   \\ 
& Flan-T5-large-770M    & 0.65      & 0.64   & 0.64   \\ 
& \textbf{GPT-2}   & \textbf{0.70}      & \textbf{0.71}   & \textbf{0.70}  \\ \hline
\end{tabular}
}
\caption{Evaluation of LLMMs with various unimodal language models as backbones and laboratory values for classifying multiple diseases.}
\label{tab:multiclass}
\end{table}

\section{System Design}

\subsection{Patient Query System}

The overall system design is depicted in Figure \ref{fig:system_design}. Our web application boasts a React-built frontend hosted on AWS EC2 and deployed using Docker containers. The user interface consists of five distinct pages: 1. Login portal, 2. Patient record management, 3. Chronic disease prediction, 4. Potential chronic disease risk alert, and 5. Early diabetes prediction.

The back end of our system includes three components. The primary component utilizes the Django MVC framework, which is deployed on AWS with Docker. This segment manages all API requests from the front end and defines the schema for the PostgreSQL database, which is hosted on a separate virtual machine. Additionally, serverless endpoints are established for necessary LLMM computations.

During patient data entry, the front end transmits it to the Django server. The back end takes charge, utilizing an API to forward the data to the LLMMs endpoint for processing asynchronously. Simultaneously, the back-end organizes and uploads the data to the database server. Once processing is complete, the back-end retrieves the processed data from the LLMMs endpoint and stores it in the database. When the front end requests specific patient predictions, the back end fetches the processed data in real-time and delivers it back to the user interface.

In addition to developing a web-based front-end, we also created a six-page Android application. The mobile application includes five user interface pages and one patient list page. This patient list page enables physicians to directly access and select patients for whom they have previously entered data. 
We will further describe our platform details in subsection \ref{sec:Demonstration_section}.

\subsection{Demonstration}\label{sec:Demonstration_section}

\subsubsection{Medical Diagnostic Web Platform}

In our chronic disease prediction platform, we demonstrate a web interface that can retrieve historical EHRs of three diseases based on the patient ID from PostgreSQL. The retrieved data can be synchronized and sent to the backend, where trained LLMMs return the risk probabilities for three disease risks to the frontend UI interface, as shown in Figure \ref{fig:web_UI} (a).

\subsubsection{New-Onset Diabetes Prediction System}

To address the clinically early diabetes risk, we designed the platform to return predictions of early diabetes risk according to different new-onset disease days from LLMMs to the front. As shown in Figure \ref{fig:web_UI} (b), for diabetes prediction, we provide early risk predictions for 90, 180, 270, and 360 days. To better visualize risk in web interface, we created a bar chart that shows the probabilities of diabetes occurrence at different periods based on the EHRs of a single patient with diabetes. 

\subsubsection{Explainable EHR Risk Assessment}

Specifically, our LLMMs framework can effectively use Shapley Additive exPlanations (SHAP) values \cite{lundberg2017unified} to highlight the risk levels in clinical notes as shown in Figure \ref{fig:web_UI} (c). The SHAP was developed from game theory and generate Shapley values for explaining the importance of features.  We first pre-train the LLMMs to focus on word position during encoding, enabling the calculation of attention scores. Subsequently, we utilize SHAP values to analyze the combined corpus of clinical notes and textual laboratory data. This visualization tool helps us understand the individual contributions of words within the corpus. By highlighting each word's positive or negative influence on predicting specific clinical terms from the LLMMs output, SHAP values enhance the model's clinical interpretability.

Finally, as illustrated in Figure \ref{fig:web_UI} (d) of our presented web interface, we designed a allowing physicians to freely upload patients' blood test items of health records and send them back to the server.  Our model can then perform modality predictions based on the same patient's record.

\subsubsection{Mobile Platform}

In this work, we designed a flexible mobile patient query system platform that can synchronize with the server and update with the backend server. As shown in Figure \ref{fig:mobile_UI} (a), the mobile interface provides doctors with a list of patients' appointments throughout the day. Subsequently, in Figure \ref{fig:mobile_UI} (b), historical clinical notes can be retrieved in real time using the patient ID for LLMMs inference. Figures \ref{fig:mobile_UI} (c) and (d) demonstrate the real-time of predictions with the web interface. Additionally, Figure \ref{fig:mobile_UI} (e) also shows how real-time blood test data can be submitted to the mobile app. This data is then synchronized with the web interface and used to update the database.

\section{Conclusion}

In this paper, we propose a medical diagnosis system platform that integrates multimodal data into LLMMs with Flask and PostgreSQL. This allows us to update patient EHR information more effectively on web and mobile platforms in real time to implement a chronic disease alert system. Furthermore, our system in the future can also provide users and clinicians with an interactive platform and disease prevention effects.
\bibliographystyle{ACM-Reference-Format}
\bibliography{references.bib}

\end{document}